# A Discrete Choice Framework for Modeling and Forecasting The Adoption and Diffusion of New Transportation Services

April 27, 2017

**Feras El Zarwi** (corresponding author)
Department of Civil and Environmental Engineering
University of California at Berkeley
116 McLaughlin Hall
Berkeley, CA 94720-1720
feraselzarwi@gmail.com
+1 510 529 8357

**Akshay Vij**
Institute for Choice
University of South Australia
Level 13, 140 Arthur Street
North Sydney, NSW 2060
vij.akshay@gmail.com
+61 08 8302 0817

**Joan L. Walker**
Department of Civil and Environmental Engineering
University of California at Berkeley
111 McLaughlin Hall
Berkeley, CA 94720-1720
joanwalker@berkeley.edu
+1 510 642 6897

## ABSTRACT

Major technological and infrastructural changes over the next decades, such as the introduction of autonomous vehicles, implementation of mileage-based fees, carsharing and ridesharing are expected to have a profound impact on lifestyles and travel behavior. Current travel demand models are unable to predict long-range trends in travel behavior as they do not entail a mechanism that projects membership and market share of new modes of transport (Uber, Lyft, etc). We propose integrating discrete choice and technology adoption models to address the aforementioned issue. In order to do so, we build on the formulation of discrete mixture models and specifically Latent Class Choice Models (LCCMs), which were integrated with a network effect model. The network effect model quantifies the impact of the spatial/network effect of the new technology on the utility of adoption. We adopted a confirmatory approach to estimating our dynamic LCCM based on findings from the technology diffusion literature that focus on defining two distinct types of adopters: innovator/early adopters and imitators. LCCMs allow for heterogeneity in the utility of adoption for the various market segments i.e. innovators/early adopters, imitators and non-adopters. We make use of revealed preference (RP) time series data from a one-way carsharing system in a major city in the United States to estimate model parameters. The data entails a complete set of member enrollment for the carsharing service for a time period of 2.5 years after being launched. Consistent with the technology diffusion literature, our model identifies three latent classes whose utility of adoption have a well-defined set of preferences that are significant and behaviorally consistent. The technology adoption model predicts the probability that a certain individual will adopt the service at a certain time period, and is explained by social influences, network effect, socio-demographics and level-of-service attributes. Finally, the model was calibrated and then used to forecast adoption of the carsharing system for potential investment strategy scenarios. A couple of takeaways from the adoption forecasts were: (1) placing a new station/pod for the carsharing system outside a major technology firm induces the highest expected increase in the monthly number of adopters; and (2) no significant difference in the expected number of monthly adopters for the downtown region will exist between having a station or on-street parking.

## 1. INTRODUCTION

The growth in population and urban development has impacted societies in one way or another from air pollution to greenhouse gas emission, climate change and traffic congestion. This made policy makers more inclined towards the development of smart cities that promote sustainable mobility, connectivity and multimodality. As such, major technological and infrastructural changes are expected to occur over the next decades such as the introduction of autonomous vehicles, advances in information and communication technology, California high speed rail, carsharing and ridesharing. This will induce potential paradigm shifts in the cost, speed, safety, convenience and reliability of travel. Together, they are expected to influence both short-term travel and activity decisions, such as where to go and what mode of travel to use, and more long-term travel and activity decisions, such as where to live and how many cars to own. This transformative mobility, whether in the form of sharing economy, connected vehicles, autonomous and app-driven on-demand vehicles and services will revolutionize travel and activity behavior.

Travel demand models are the commonly-used approach by metropolitan planning agencies to predict 20-30 year forecasts of traffic volumes, transit ridership, walking and biking market shares across transportation networks. These models try to assess the impacts of transportation investments, land use and socio-demographic changes on travel behavior with the main objective of predicting future mode shares, auto ownership levels, etc. These models focus on a behaviorally richer approach to modeling travel mode choice as opposed to the traditional four step travel demand models. Travel demand models evaluate travel and activity behavior as a series of interdependent logit and nested logit models that entail travel mode choice, vehicle availability, and time-of-day models, etc. However, current travel demand models are unable to predict long-range trends in travel behavior as they do not entail a mechanism that projects membership and market share of new modes of transport (Uber, Lyft, autonomous vehicles, etc). According to Guerra (2015), "only two metropolitan planning organizations in the 25 largest metropolitan areas mention autonomous or connected vehicles in their long-range regional transportation plans". That is why current travel demand models lack a methodological framework that caters for those upcoming transportation services and technologies and their impact on travel behavior which will be prevalent in 20-30 years.

Our objective is to develop a methodological framework tailored to model the technology diffusion process by focusing on quantifying the effect of the spatial configuration of the new technology and socio-demographic variables. Moreover, we are also interested in capturing the effect of social influences and level-of-service attributes of the new technology on the adoption process. The methodological framework used in our analysis entailed an integrated latent class choice model (LCCM) and network effect model that was governed by a destination choice model. Our approach was confirmatory as the latent classes used in the analysis (innovators/early adopters, imitators and non-adopters) are rooted in the technology diffusion literature across multiple disciplines. These latent classes are able to capture heterogeneity in preferences towards technology adoption. Our research is motivated by existing work in technology adoption modeling which employs a microeconomic utility-maximizing representation of individuals. This framework is of interest to us as it could be easily integrated with our disaggregate activity-based models. Our proposed disaggregate technology adoption model shall help planners and policy makers gain insight regarding the projected market shares of upcoming modes of transport for various policies and investment strategies at the public and private levels.

Most diffusion models employ an aggregate framework, for example the Bass model (Bass, 1969). While recent aggregate models have further enriched the specification of the Bass model, they still do not account for a range of policy variables (including the spatial configuration) that can be used to rank policies and investment strategies needed to maximize the expected number of adopters of a new technology in future time periods. Our methodological framework is different than other disaggregate models in the diffusion and transportation literature as it accounts for (1) heterogeneity in the decision-making process across distinct market segments that have a different adoption behavior; and (2) the spatial configuration effect of the new technology in terms of quantifying how an increase in the size of the network governed by the new technology will impact adoption.

This study contributes to the existing body of literature in providing a unique methodology to model the adoption behavior and uptake of new products/technologies by various market segments. Our model caters for the effects of social influences, network effect, socio-demographics and level-of-service attributes of the product on the adoption behavior of each of the market segments. The following framework could be used to predict future market shares of upcoming modes of transport as one specific type of application. The paper is organized as follows: Section 2 provides a literature review of existing technology adoption and diffusion models. Section 3 provides the adopted methodological framework used to model technology adoption and details the framework of the dynamic Latent Class Choice Model (LCCM) and the network effect model. Section 4 explains the dataset used in the study. Section 5 discusses model results and model applications. Section 6 focuses on comparing forecasts between our proposed generalized adoption model and the Bass aggregate diffusion model for three different policy scenarios. Section 7 concludes the findings of the paper.

## 2. LITERATURE REVIEW

Autonomous vehicles are on the horizon, not to mention the transformative mobility trend that is occurring in our transportation system via the introduction of electric vehicles, ridesharing, carsharing, and many other new technologies. In order to quantify the effect of transportation policies and investment strategies in a representative manner, travel demand models should be able to model and forecast market shares for those new modes of transportation. However, current travel demand models do not entail a mechanism to do so, which in turn provides the core motivation behind this paper. We believe that models of technology adoption and diffusion, which are widely used across multiple industries and cultures to forecast uptake of new technologies, will bridge this gap in the transportation literature. Diffusion models are popular in a variety of disciplines such as: agriculture (Sunding and Zilberman, 2001; and Ward & Pulido-Velazquez, 2008), consumer durables (Delre et al., 2007; and Schramm et al., 2010), pharmaceutical industry (Desiraju et al., 2004), and the automobile industry and in particular aggregate diffusion patterns of car ownership (Dargay and Gately, 1999). Also, diffusion models have been estimated and used in forecasting across different cultures such as: United States, France, Spain and many other countries (please refer to Tellis et al., 2003).

Over the course of the next few paragraphs we will describe the central piece of the framework governed by the model of technology adoption. The adoption and diffusion of new technologies have received attention across multiple disciplines within economics and social sciences over the years. As defined by Rogers (1962), "diffusion is the process by which an innovation is communicated through certain channels over time among the members of a certain social system". Any innovation may be defined in terms of the relative advantage offered by the innovation over existing alternatives, the degree to which the innovation is consistent with existing needs and values, the measure of difficulty associated with using the innovation, the extent to which the innovation can be tried on a limited basis, and the ease with which the benefits of the innovation are tangible to others. Differences in social systems may be characterized by the pattern of relationships among members of the system, established norms of what constitutes acceptable and unacceptable behavior, and the degree to which individual agents are able to influence the behavior of others. Communication channels can be broadly classified as either mass media, such as the television, or interpersonal channels that require a direct exchange between two or more individuals.

Diffusion models are widely used in the marketing science domain and many other industries, as mentioned above, as they capture the dynamics behind the uptake of a new product in addition to forecasting its demand. Forecasting accuracy with diffusion models varies depending on the type of dataset being used, whether it's homogenous or heterogeneous i.e. from different sources (Meade and Islam, 2006).

Improvement with respect to specification of the diffusion models such as incorporating non-parametric parametrization and enhancing flexibility has helped increase forecasting accuracy across multiple disciplines (Meade and Islam, 2006).

Rogers (1962) defines the following five classes of adopters that influence the uptake of a certain technology across various disciplines: innovators, early adopters, early majority, late majority and laggards. Based on the mathematical formulation of the diffusion model of Bass (1969), adopters can be divided into two distinct groups: innovators and imitators with the latter comprising the remaining four classes of adopters listed above. The technology diffusion literature stresses on the importance of the role of those two different types of adopters in shaping the market penetration rate of a new good or service (please refer to Mansfield, 1961; Mahajan et al., 1990; and Cavusoglu et al, 2010). Innovators are individuals that "decide to adopt an innovation independently of the decisions of other individuals in a social system" while imitators are adopters that "are influenced in the timing of adoption by the pressures of the social system" (Bass, 1969).

Throughout the next few paragraphs, we will describe the assumptions and formulations of aggregate and disaggregate technology adoption models, and motivate why disaggregate models are a better methodological approach to our research question. Aggregate models of technology diffusion formulate the percentage of the total population that has adopted an innovation at some time period as some function of the characteristics of the population and the attributes of the innovation. The empirical research on aggregate models was pioneered by Griliches (1957), Mansfield (1961), and Bass (1969).The Bass model is well-known in the marketing science literature and it formulates the probability that a certain consumer will make an initial purchase at a given time t given that no purchase has been yet made by that specific consumer denoted as $P_t$ in the equation below as a linear function of the number of previous buyers:

$$P_t = p + \frac{q}{M} Y(t)$$

$p$: Coefficient of innovation; $q$: Coefficient of imitation; $M$: Total potential market for the technology

$Y(t)$: Cumulative number of individuals that adopted the new technology by time t (number of previous buyers)

The term $\frac{q}{M} Y(t)$ reflects the "pressures operating on imitators with an increase in the number of previous buyers" (Bass, 1969) while $p$ reflects the percentage of adopters that are innovators.

Using this formulation, sales of a certain technology/product could be forecasted into the future via a closed form solution. We are interested in the formulation of the Bass model as it identifies the two types of adopters of a new technology in addition to capturing the effect of social influence onto the probability of adoption. The figure below depicts the sales of a product over time (bell-shaped curve, S(t)) and cumulative sales over time ("S"-shaped curve, Y(t)) according to Bass (1969). The plot below uses a value of 0.005 for the coefficient of innovation p, 0.3 for coefficient of imitation q, and 100 for total potential market M. Those values were chosen arbitrarily to display the shape of the S(t) and Y(t) curves and provide useful insights. It is evident from the "S"-shaped diffusion curve that once a certain good or service is introduced in a market, it exhibits a low adoption rate followed by takeoff whereby the market experiences high adoption rates. After the takeoff period, technology adoption slows down until it reaches market saturation.

Mansfield (1961) on the other hand formulates the cumulative sales of a good/service using a logistic model, which is a special case of the Bass model (p=0). Extensions of the Bass model and more recent enhancements to aggregate diffusion models (see for example Kamakura and Balasubramanian, 1988; and Meade & Islam, 2006), have incorporated the effects of price, advertising and other marketing variables into the model parametrization in an attempt to increase forecasting power. Furthermore, aggregate models have been developed to assess the diffusion levels of a certain technology across different countries.

Recently, agent-based modeling and simulation methods are becoming more popular in the technology diffusion discipline as they are estimated on an individual level. This will in turn address some of the shortcomings of aggregate diffusion models and cater for heterogeneity among consumers and explicit social structure (Kiesling et al., 2012; and Schramm et al., 2010).

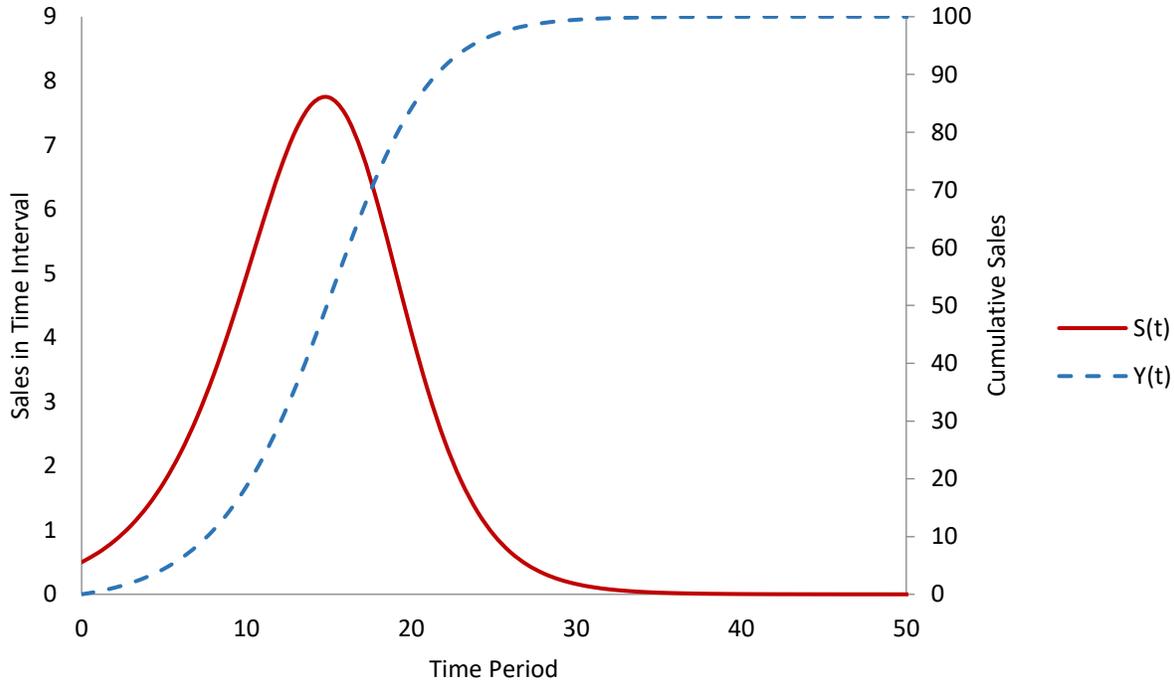

**Figure 1: Sales vs. Cumulative Sales over Time**

Disaggregate models of technology adoption on the other hand formulate the probability that an individual or household adopts an innovation as some function of the characteristics of the decision-maker, attributes of the alternative, communication channels (both interpersonal networks and mass media) and time in order to cater for the temporal dimension of the diffusion process. These models have been used to predict the adoption of a wide variety of technologies and innovations that include color televisions, genetically modified crops, irrigation technology, computers, diapers and drill bits (Zilberman to al., 2012). Disaggregate models are of interest to us for the following reasons: (1) they employ a microeconomic utility-maximizing representation of individuals that provides insight into the decision-making process underlying the adoption or non-adoption of different innovations by consumers, which is consistent with the framework typically employed by travel demand models; (2) they capture various sources of heterogeneity in the decision-making process that will drive different consumers to adopt at different times; (3) they can be transferable across different geographical, social and cultural contexts with pertinent model calibration; and (4) they can account for a range of policy variables that can be used to rank policies and investment strategies in terms of maximizing the expected number of adopters of a new technology in future time periods. Moreover, we are interested in understanding how the spatial configuration of a new transportation service and the different socio-demographic variables of decision-makers can influence the adoption behavior. The aforementioned aggregate models cannot cater for those two key variables in their formulation to project future market shares of a new technology in a more representative manner. In addition to that, model application is a key component in our analysis as it provides policy makers and transportation specialists with the means to quantify the expected number of adopters for a set of policies and strategies at the metropolitan levels. Aggregate models do suffer from a limited degree of policy sensitivity and can only account for a narrow range of policy variables which make them less appealing to our analysis.

There are various disaggregate diffusion models, each focusing on different aspects of the decision making process and behavior. One dominant disaggregate adoption model is the threshold model which was first introduced by David (1969) in an attempt to study the technology adoption of grain harvesting machinery and was further explored by Sunding and Zilberman (2001). The threshold model incorporates heterogeneity among decision-makers in the adoption process and could be used in conjunction with discrete choice models (logit or probit) to represent the utility maximization behavior of decision-makers. The sources of heterogeneity that affect the adoption process may include various variables depending on the available data and what the analyst is trying to capture. At every time period, the critical level of each source of heterogeneity in the model is determined. Decision-makers equipped with a value of that source of heterogeneity, say income, that is larger than the critical level at a certain time period will choose to adopt the new technology/product at that time period. The critical level of a source of heterogeneity shall decrease over time which induces more consumers to adopt due to principles of "learning by doing" and "learning by using" (please refer to Sunding and Zilberman, 2001). One application of this consisted of using a disaggregate utility function model of household vehicle choice using the threshold model in its aggregate context with income, household structure, comfort/quality being three critical sources of heterogeneity (Liu, 2010). Advances in the threshold model incorporate dynamic optimization in their analysis, such that a decision-maker is making a trade-off between the expected decrease in price of a certain technology in the future and the current benefits from purchasing it which will dictate the timing of adoption (McWilliams and Zilberman, 1996).

As we are interested in capturing various sources of heterogeneity in the decision-making process, the threshold model does not seem to be a good fit to the methodological framework we want to adopt. As previously mentioned, the literature focuses on two different types of adopters (early adopters and imitators). We are interested in modeling the adoption behavior of those two distinct market segments in addition to the non-adopters market segment that chooses to never adopt a new technology. The formulation of the disaggregate utility function of the threshold model can be used as a starting point in the development of our methodological framework of technology adoption for the three different market segments.

What about the transportation industry? The transportation industry has been trying to develop quantitative methods rooted in the technology diffusion literature to try and predict market shares of those upcoming modes of transportation. One study (Li et al., 2015), focused on defining variables that influence ridership of the Taiwan High Speed Rail System (THSR) using econometric time series models and revealed preference (RP) data of monthly ridership from January 2007 till December 2013. Two models were estimated: (1) seasonal autoregressive integrated moving average and (2) first order moving average model to explore the influence of explanatory variables on ridership.

Moreover, studying the market diffusion of electric vehicles has received worldwide attention these past few years. For example, Plötz et al. (2014), estimated an agent-based simulation model of the diffusion process of electric vehicles using real-world driving data that captured heterogeneity among decision-makers, psychological factors and attributes of the new technology. Another study, please refer to Gnann et al. (2015), used an Alternative Automobiles Diffusion and Infrastructure (ALADIN) diffusion model to forecast market penetration of plug-in electric vehicles through simulation techniques. Their proposed methodology incorporated an agent-based simulation model that catered for different types of users in addition to their respective decision making processes to make it behaviorally richer. Other studies focused on using agent-based simulation models alone while others integrated them with discrete choice methods to account for a richer behavioral interpretation (Eppstein et al., 2011; Brown, 2013; Zhang et al, 2011). For example in Eppstein et al. (2011), an integrated agent-based and consumer choice model was estimated that tried to capture the effect of social interactions and media on the market penetration of plug-in hybrid electric vehicles.

The adoption of new transportation services has primarily focused on using stated preference (SP) data in the context of alternative fuel vehicles. Some studies were interested in assessing sensitivities to attributes of the new technology (Ito et al., 2013 and Hirdue et al., 2011) while other studies focused on both model estimation and forecasting the market share of alternative fuel vehicles under certain policy scenarios (Glerum et al., 2013, & Mabit and Fosgerau, 2011). The SP approach does capture sensitivities to attributes of the new technology in a representative way. However, using SP data does require solid model calibration and validation to enhance the model's forecasting power. In order to account for this, integrating SP with revealed preference (RP) data would be a better approach (see for example Brownstone et al., 2000). Ideally, one should be interested in using RP data as it represents actual market demand. An SP approach entails hypothetical scenarios which hinders a model's forecasting power. In addition to that, the analyst will not be able to capture the dynamic aspect of the diffusion process over time with respect to the social influence and spatial component dimensions of the new technology. In previous SP studies, projected market shares for electric vehicles and alternative fuel vehicles were over estimated. That is due to the fact that the adoption and diffusion of a new technology is a temporal and social process and these previous studies did not account for this.

Also, a current developed model focuses on forecasting adoption of electric vehicles using an integrated discrete choice and diffusion models (Jensen et al., 2016). This model builds on the previous work of Jun and Park (1999) whereby they specify the utility of adopting a certain good at time t as a function of the attributes of the technology, and difference between time t and the time period at which the product was introduced in the market. The parameter associated with the aforementioned second variable in the utility of adoption will account for the effect of the diffusion process. The probability of adoption at a certain time period could be computed using the logit closed form. Following that, the sales of electric vehicles at different time periods could be computed respectively. To forecast the demand of electric vehicles, data was collected from a stated preference (SP) survey conducted in Denmark in 2012 and 2013 for the choice between electric vehicles and internal combustion engines. The specification of the utility of choosing either mode included purchase price, propulsion costs, driving range, emissions, number of battery stations, and characteristics of public charging facilities. The utility equation of choosing an electric vehicle also entailed a parameter that portrays the effect of the diffusion process while assuming that internal combustion engines have reached market saturation. The model was used to forecast market share of electric vehicles for several policy scenarios. Our proposed methodological framework is different as it caters for (1) heterogeneity among decision-makers and in particular among distinct discrete market segments in the population that have different adoption behavior; (2) effect of various socio-economic and demographic variables on the diffusion process; (3) spatial or network effect of the new technology whereby we are interested in assessing how an increase in the size of the network that is covered by the new mode of transportation will impact adoption behavior; and (4) social influences and how that will influence the utility of adoption.

## 3.  METHODOLOGICAL FRAMEWORK

The methodological framework we want to develop builds on the aggregate diffusion literature and in particular the concepts of consumer heterogeneity towards the adoption process i.e. innovators versus imitators, and social influences as described in the Bass model. We are interested in disaggregate diffusion models as they can be easily integrated with the activity-based travel demand models of interest. Also, with disaggregate models, we can account for the impact of socio-demographics and social influences on the adoption process in addition to spatial effects. By spatial effects we are referring to increasing the relative size of potential destinations that one can reach out to via the new mode of transport. While there have been disaggregate models developed in the literature, they seem to be based on different behavioral assumptions (for example the previously mentioned threshold model) or do not cater for heterogeneity in the specification of the utility of adoption. Most studies in the literature focus on the role of three defined distinct market segments in their analysis that differ in their respective adoption behavior towards a new technology. Those market segments are: innovators/ early adopters, imitators and non-adopters. We will be building on these findings using a disaggregate technology diffusion approach, which is rooted in findings from the aggregate diffusion literature.

The specification we are interested in developing is unique as it tries to model how technology adoption and use is influenced by socio-demographics, attributes of the new technology/service, spatial effect (or network effect) and finally social influences. The aggregate diffusion literature mainly refers to two types of adopters (innovators and imitators). In order to assess the adoption behavior of a certain population we need to take into account those decision-makers that will choose to never adopt the new technology/service. We are interested in modeling the adoption behavior of each of the following three market segments (innovators/early adopters, imitators, and non-adopters) to try and capture heterogeneity in the adoption behavior of each of those market segments. Innovators or early adopters denote the market segment that determines whether a new technology will pick up in market share or not after being introduced in the market. They define how steep or flat the "S" cumulative diffusion curve can be during the early stages. Innovators comprise the biggest fraction of adopters of a new technology during the initial time periods. Imitators on the other hand come into play as time elapses since the introduction of the new technology. They will determine the rate at which the market will adopt the new product or service and will in turn shape the steepness of the "S" cumulative diffusion curve at later stages in the diffusion process. Non-adopters will define the time period at which the cumulative diffusion curve reaches a plateau. For example, as the number of non-adopters increases the faster the "S" curve attains a plateau.

However, we do not observe what type of a person any given individual is i.e. we do not have information about which market segment each decision-maker belongs to. In order to account for this, discrete mixture models and in particular latent class choice models (LCCM) are found to be the most appropriate framework. Latent class choice models comprise two components: a class membership and a class-specific choice model as depicted in the figure below.

The class-specific choice model formulates the probability of technology adoption of a certain individual conditional on that individual either being an innovator, imitator or non-adopter. This component captures variation across classes with respect to choice set, tastes and sensitivities, decision protocol and covariance structure of the error term (Gopinath, 1995).

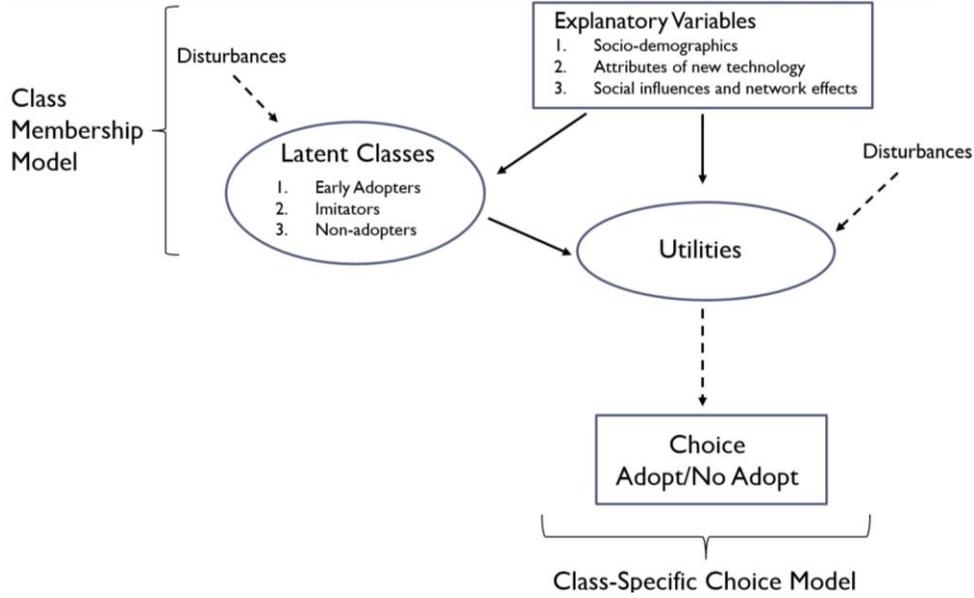

**Figure 2: Latent Class Choice Model Framework**

As we are interested in modeling the adoption process for each market segment, we should cater for the temporal dimension of technology diffusion as decision-makers will adopt the new technology at various time periods according to the aforementioned explanatory variables. Hence, the probability that individual n during time period t after the new technology was available in the market adopted or chose to not adopt could be written as:

$$P\left(y_{ntj}\middle|Z_{nt}, X_{ntj}, q_{ns}\right) \forall j \in \{0,1|y_{n(t-1)j}\}$$

where $y_{ntj}$ equals one if individual n during time period t chose to adopt the new technology (j=1) and zero otherwise, conditional on the characteristics of the decision-maker during time period t denoted as $Z_{nt}$ and attributes of the new technology (j=1) during time period t denoted as $X_{ntj}$, and conditional on the decision-maker belonging to latent class s ($q_{ns}$ equals one and zero otherwise).

Now, evaluating the probability of adoption or non-adoption will be based on a binary logit formulation that transforms the utility specification into probabilities. Let $U_{ntj|s}$ denote the utility of adoption (j=1) or not (j=0) of the new technology during time period t for individual n conditional on him/her belonging to latent class s which is expressed as follows:

$$U_{ntj|s} = V_{ntj|s} + \varepsilon_{ntj|s} = z'_{nt}\beta_s + x'_{ntj}\gamma_s + \varepsilon_{ntj|s}$$

where $V_{ntj|s}$ is the systematic utility that is observed by the analyst, $z'_{nt}$ is a row vector of characteristics of the decision-maker n during time period t, $x'_{ntj}$ is a row vector of attributes of the new technology (j=1) during time period t for individual n, $\beta_s$ and $\gamma_s$ are column vectors of parameters specific to latent class s and $\varepsilon_{ntj|s}$ is the stochastic component of the utility specification. Since we have prior assumptions about the behavior of the two various types of adopters (innovators versus imitators) based on the existing technology diffusion literature, the systematic utility of adoption for each of the three latent classes was specified according to the following rationale. The systematic utility of adoption of innovators shall include characteristics of the decision-maker and attributes of the new technology as we are interested in assessing the significance of those explanatory variables on the decision process of adopting or not. The systematic utility of adoption for imitators is also modeled as a function of the characteristics of the decision-maker and attributes of the new technology. However, this is the latent class whose adoption behavior is influenced by the extent of social influence and accumulating pressure with the increase in the previous number of adopters (Bass, 1969). That is why we are interested in determining the effect of the previous number of adopters on the utility of adoption of imitators at a certain time period. Finally, the systematic utility of adoption of the third latent class (non-adopters) consists of an alternative specific constant (ASC) only. Ideally, this ASC should attain a highly negative value via estimation to ensure that this class will most likely never adopt the new technology. The systematic utility of adoption / non-adoption for innovators, imitators and non-adopters is specified in the following manner:

$$\begin{cases} V_{adopt,n,t|s=innovator} = z'_{nt}\beta_1 + x'_{ntj}\gamma_1 \\ \qquad V_{non-adopt,n,t|s=innovator} = 0 \end{cases}$$

$$\begin{cases} V_{adopt,n,t|s=imitator} = z'_{nt}\beta_2 + x'_{ntj}\gamma_2 + \Delta_{(t-1)}\alpha_2 \\ \qquad V_{non-adopt,n,t|s=imitator} = 0 \end{cases}$$

$$\begin{cases} V_{adopt,n,t|s=non-adopter} = \lambda \\ V_{non-adopt,n,t|s=non-adopter} = 0 \end{cases}$$

where $\Delta_{(t-1)}$ depicts the cumulative number of adopters of the new technology during time period (t-1), and $\lambda$ is an alternative specific constant.

Now, in order to assess the impact of the spatial/network effect of the new technology on the utility of adoption, we were interested in quantifying the level of accessibility brought about by the new mode of transportation. Accessibility is defined as the "ease with which any land-use activity can be reached from a location, using a particular transport system" (Dalvi et al., 1976). There are several types of accessibility measures: cumulative opportunities measures, gravity-based measures, and utility-based measures (Handy and Niemeier, 1997). We will focus on utility-based measures for the assessment of accessibility through developing a destination choice zone-based model. Utility based measures of accessibility have desirable advantages over other methods as they account for flexibility in travel purposes and sensitivity to travel impedance measures in terms of time and cost. Also, they capture individual-level preferences and socio-demographic influences on travel behavior. In those types of models, we assume that given a certain origin, each decision-maker associates a utility to each of the available destinations in his/her respective choice set $C_n$ and will end up choosing the alternative i.e. destination which maximizes his/her utility. Accessibility is defined as the logsum measure of those destination choice models as it "measures the expected worth of certain travel alternatives" (Ben-Akiva and Lerman, 1985).

Let $U_{nij}$ denote the utility of individual n conducting a trip from origin i to destination alternative j. Determining the systematic utility specification requires assessing the explanatory variables that influence an individual's decision to conduct a trip from a certain origin to a certain destination. Travel impedance whether in terms of travel distance or cost is an important variable as travelers prefer conducting shorter trips. Second, since travel is a derived demand whereby an individual goes from a certain origin to a destination to conduct an activity, evaluating the available number of opportunities or attractions at the destination is important. In addition to that, an individual is more likely to use the new technology (mode of transport in our case) if it provides a relatively close destination spot to his/her home. Finally, socio-demographic variables can play a role in defining some characteristics that can drive individuals into conducting certain trips. Accordingly, $U_{nij}$ was specified in the following manner:

$$U_{nij} = V_{nij} + \varepsilon_{nij} = d_{ij}\beta + \ln(size_j)\alpha + Z_n\gamma + X_{nj}\theta + home_n\delta + \varepsilon_{nij}$$

where $V_{nij}$ is the systematic utility observed by the analyst, $d_{ij}$ denotes a friction factor of traveling from origin i to destination alternative j which is the travel distance associated with origin-destination pair (i,j), $size_j$ represents the attractions associated with destination j which will be governed by the employment rate at the destination (number of employees per square mile) as it is considered to be the driver behind trip attractions, $Z_n$ represents socio-demographic characteristics of decision-maker n, $X_{nj}$ denotes attributes of the new technology at destination alternative j for individual n, $home_n$ is a dummy variable which will be equal to one if decision-maker n resides within a certain proximity from his/her corresponding destination alternative and zero otherwise, $\beta$, $\alpha$, $\gamma$, $\theta$, and $\delta$ are parameters associated with the explanatory variables, and $\varepsilon_{nij}$ is the stochastic component of the utility specification.

Assuming that all individuals are utility maximizers and that $\varepsilon_{nij}$ follows an i.i.d. Extreme Value Type I distribution across individuals, origin and destination alternatives with mean zero and variance $\frac{\pi^2}{6}$, the accessibility measure is expressed as the following logsum measure:

$$Accessibility_{n,i,t} = ln\left[\sum_{j=1}^{J_t} e^{V_{nij}}\right]$$

where i denotes an origin alternative and $J_t$ is the total number of distinct destination alternatives available at time period t.

Accessibility changes over time due to an increase/decrease in the number of distinct destination alternatives $J_t$ or changes in any of the explanatory variables of the destination choice model systematic utility. Changes in the employment rate, socio-demographics, or attributes of the new technology will induce changes in the accessibility measure over time.

Based on the above formulation, the difference in the utility of adoption for the two types of adopters (early adopters and imitators) comprises different sensitivities to characteristics of the decision-maker and attributes of the new technology. In addition to that, the adoption process for imitators is affected by the social influence aspect of the new technology while early adopters are not. Let us focus on a carshring service as an example. Assume a decision-maker resides in a certain area where the accessibility measure of the new technology is unattractive mainly because the individual resides in an area that is far away from zones that have a station for this carsharing service. Also, the nearest station for this service is relatively far from the decision-maker's home. In this case, the attributes of the new technology are undesirable, and that individual is unlikely to adopt. However, as the technology evolves and becomes more attractive, that individual is more likely to adopt. This decision-maker is an early adopter in his/her local context even though he/she adopted at a later point in time. Our methodological framework caters for this as it deals with the micro-level disaggregate decision-making process. Aggregate models on the other hand, such as the Bass model, examine the diffusion process at a system level whereby they would consider this particular decision-maker to be an imitator.

Now that we have defined the formulation of the network effect model denoted by accessibility, we return to the formulation of the class-specific choice model. Assuming that all individuals are utility maximizers and that $\varepsilon_{ntj|s}$ follows an i.i.d. Extreme Value Type I distribution across individuals, time periods, alternatives and latent classes with mean zero and variance $\frac{\pi^2}{6}$, the class-specific choice model could be formulated as such:

$$P\big(y_{ntj}|Z_{nt}, X_{ntj}, q_{ns}\big) = P(U_{ntj|s} \geq U_{ntj'|s} \; \forall \, j' \in \; C) = \frac{e^{V_{ntj|s}}}{\sum_{j'=1}^{J} e^{V_{ntj'|s}}}$$

where $C$ denotes the choice set i.e. either adopting to the new service or not which is common to all individuals.

Assuming that the class-specific choice probabilities for individual n across all choice situations are conditionally independent given that he/she belongs to latent class s, then the conditional probability of observing a vector of choices $y_n$ becomes:

$$P(y_n|q_{ns}) = \prod_{t=1}^{T_n} \prod_{j \in C} P\left(y_{ntj}|Z_{nt}, X_{ntj}, q_{ns}\right)^{y_{ntj}}$$

where $T_n$ is the total number of time periods available for individual n until he /she adopts.

The class membership model on the other hand predicts the probability that decision-maker n with characteristics $Z_n$ belongs to latent class s and is defined as such:

$$P(q_{ns}|Z_n)$$

Let $U_{ns}$ denote the utility for individual n from latent class s which is expressed as follows:

$$U_{ns} = V_{ns} + \varepsilon_{ns} = z_n'\tau_s + \varepsilon_{ns}$$

where $V_{ns}$ is the systematic utility, $z_n'$ is a row vector of socio-economic and demographic variables for decision-maker n, $\tau_s$ is a column vector of parameters, and $\varepsilon_{ns}$ is the stochastic component of the utility specification. Again, assuming that all individuals are utility maximizers and that $\varepsilon_{ns}$ follows an i.i.d. Extreme Value Type I distribution across individuals and latent classes with mean zero and variance $\frac{\pi^2}{6}$, the class membership model could be formulated as such:

$$P(q_{ns}|Z_n) = P(U_{ns} \geq U_{ns'} \ \forall \ s' = 1,2,\ldots,S) = \frac{e^{V_{ns}}}{\sum_{s'=1}^{S} e^{V_{ns'}}}$$

where $S$ denotes the total number of distinct latent classes which is equal to three in our case.

Now, to put things in perspective with respect to our methodological framework, the figure below displays all three components in our analysis.

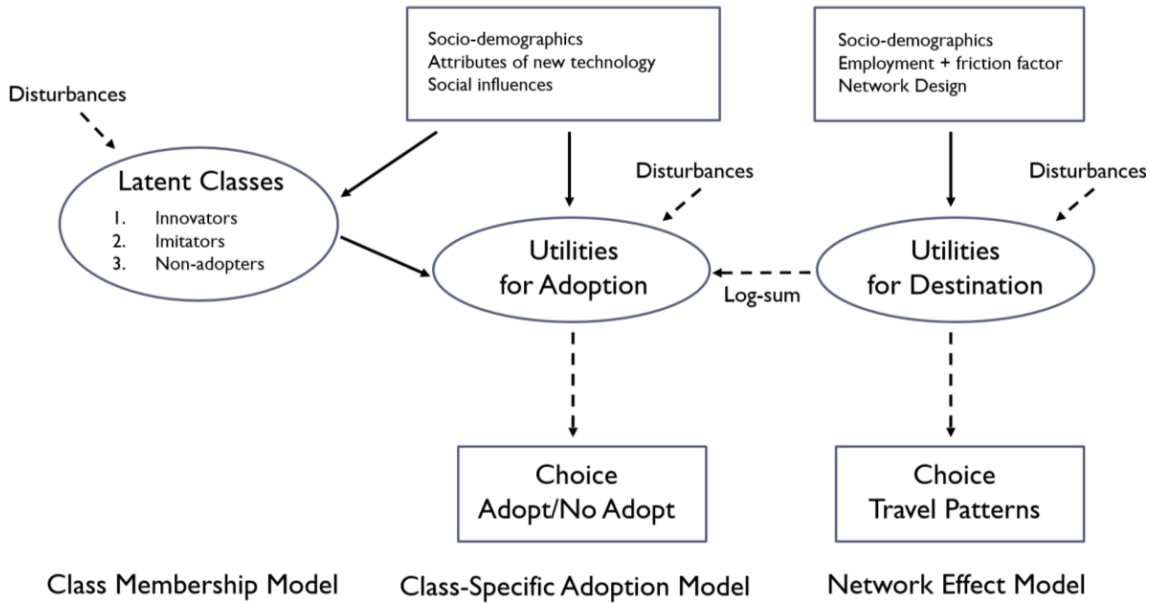

**Figure 3: Generalized Technology Adoption Model**

The destination choice model will dynamically feed into the class-specific adoption model in terms of evaluating the accessibility measure at different time periods. Afterwards, joint estimation of the class-specific adoption model and class membership model will take place.

The marginal probability $P(y)$ of observing a vector of choices y for all decision-makers is:

$$P(y) = \prod_{n=1}^{N} \sum_{s=1}^{S} P(y_n | q_{ns}) \, P(q_{ns} | Z_n) \ = \ \prod_{n=1}^{N} \sum_{s=1}^{S} P(q_{ns} | Z_n) \prod_{t=1}^{T_n} \prod_{j \in C} P(y_{ntj} | Z_{nt}, X_{ntj}, q_{ns})^{y_{ntj}}$$

Finally, the technology adoption model predicts the probability that a certain individual will adopt the new technology/service at a certain time period, and is explained by social influences, network effect, socio-demographics and level-of-service attributes. The model was estimated via the Expectation- Maximization (EM) algorithm. This optimization technique enhances the computation power of model estimation by making use of conditional independence properties that exist in our model.

## 4.  DATASET

We will use revealed preference (RP) time series data to estimate the integrated discrete choice and technology adoption model from a one-way carsharing system in a major city in the United States. The name of the carsharing company is withheld for confidentiality reasons. Our data focuses on the adopters of the service ever since it was launched. Signing up to be a member of this carsharing system requires a membership fee but no monthly nor annual fees. Currently, there are 14 pods/stations in addition to 5 locations for on-street pick-up/drop-off locations. The dataset entails zip code information about members of the new transportation service which drove our analysis to be zip code focused. In total, there are 16 zip code based stations for the car sharing service as some of the on-street pick-up/drop-off locations exist in the same zip code as other stations.

The dataset consists of all individuals that have signed up for the service for a time period of 2.5 years after being launched in addition to their registration date, gender and zip code associated with their residential location or zip code at which the registration payment was performed. Moreover, travel patterns via the carsharing service for a period of 6 months were recorded. Information about which user conducted a trip was recorded in addition to the origin and destination carsharing stations used. Our main focus revolves around the technology adoption behavior of residents of that major city and hence we are only interested in those adopters that had a location zip code affiliated with it which summed up to 1847 adopters. Initially, we had information about all members of the carsharing service but we limited our analysis to members that used the service during the last six months of the data collection period. An adopter is an individual that has signed up for the carsharing service and that has conducted at least one trip during the last six months of the data collection period. The figure below highlights the cumulative number of adopters over the entire time period that are active users of the service in order to project where exactly on the "S" diffusion curve the carsharing system's current market share is.

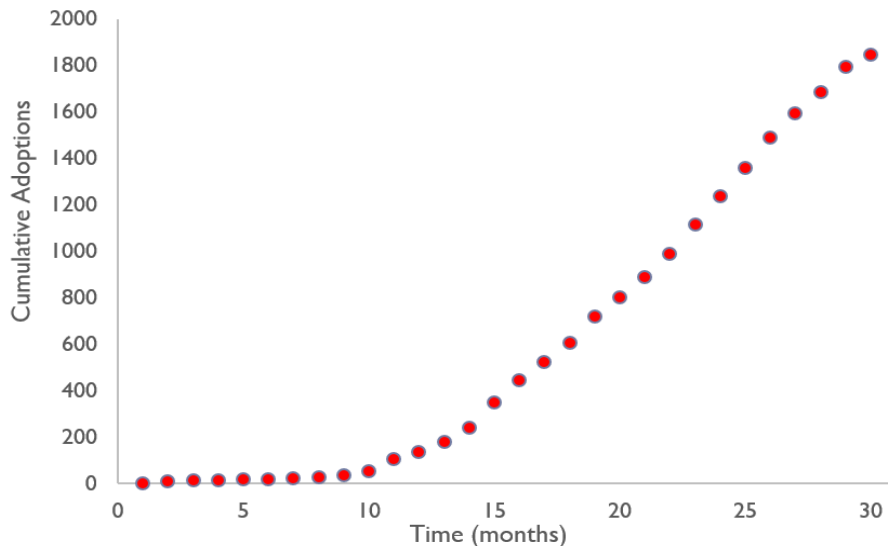

**Figure 4: Cumulative Number of Adopters of Carsharing Service**

Finally, in order to have a representative sample of the population, we wanted to enrich the sample with a random draw of 2724 observations from the Household Travel Survey (2013) of the same state to which the city we are working with belongs. We will also assume that the individuals from this random sample are non-adopters i.e. did not adopt to the new service for the entire data collection time period (2.5 years).

The prior probability of being an adopter in the city of interest is $3 \times 10^{-4}$ given the number of adopters and the population. Hence, the expected number of adopters in the random sample is approximately one.

The figure below displays the growth in the number of pods/stations and on-street pick-up/drop-off locations for the 2.5-year time period.

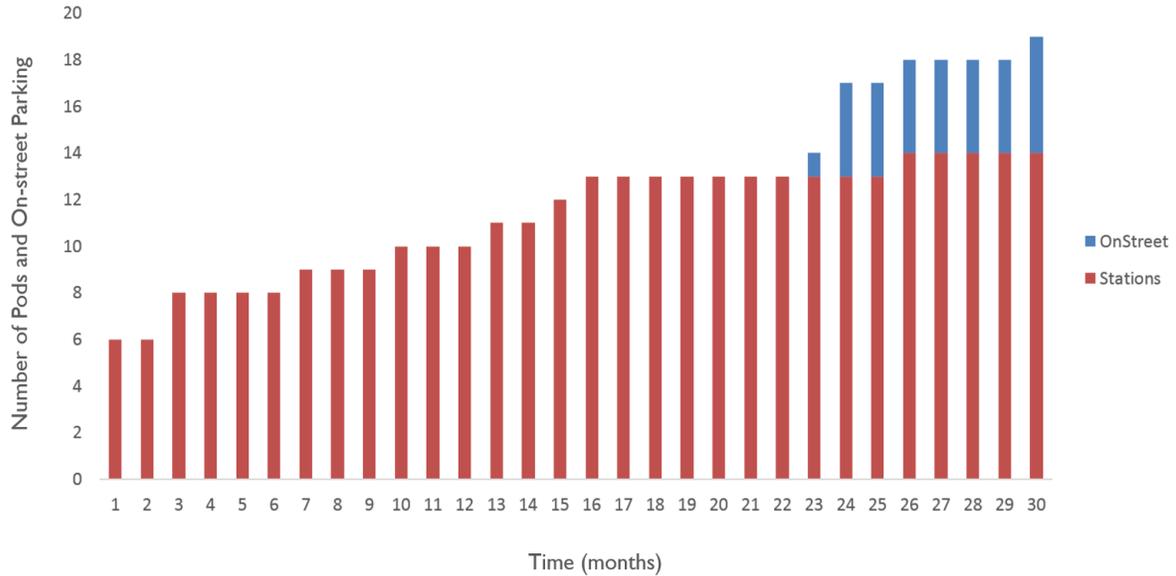

**Figure 5: Growth in Number of Pods/Stations and On-Street Parking over Time**

Our technology adoption model shall assess the impact of socio-demographics, carsharing supply (fleet and pricing), social influences and network effect on the adoption behavior of innovators, imitators and non-adopters. Identifying network effect that is governed by the construct of accessibility shall be restricted to be zip code based for the same reason mentioned above. We would like to identify the level of accessibility associated with each zip code based station of the carsharing system depending on the spatial distribution of potential destinations i.e. stations. The origins and destinations entail the full set of the carsharing system's stations. The destination choice model will be estimated based on trips that were conducted by users over a period of 6 months. For our formulation with this dataset, the accessibility measure will be non-zero only for users that have a home zip code associated with one of the stations or on-street parking locations. To account for that, we wanted to assign an accessibility measure for zip codes which do not entail a station/pod or on-street parking. We were interested in imputing the accessibility for those zip codes from the accessibility of the nearest zip code that had either a station or on-street parking while taking a friction factor into consideration, distance in our case. The accessibility measure for individual n with home zip code i which does not have a station or on-street parking at time t could be defined as follows:

$$Accessibility_{n,i,t} = \frac{Accessibility_{n,k,t| \; nearest \; active \; station \; k \; at \; time \; t}}{(distance_{i,k})^{\varphi}}$$

where $\varphi$ denotes the degree of the distance friction effect which will be estimated in the model.

Moreover, the sample population we are working is choice-based whereby each choice in the available choice set (adopt, not adopt) corresponds to a separate stratum (carsharing members versus household travel survey sample). However, the sampling fractions are not equal to the population shares especially that we have accounted for all adopters of the carsharing system and hence are highly over-represented in our sample. To cater for that and yield consistent parameter estimates, each observation needs to be weighted by $\frac{W_g}{H_g}$ where $W_g$ is the population fraction and $H_g$ is the sample fraction of members of stratum $g$ (Ben-Akiva and Lerman, 1985). Accordingly, the marginal probability $P(y)$ of observing a vector of choices for all decision-makers should be expressed as follows:

$$P(y) = \prod_{n=1}^{N} \left( \sum_{s=1}^{S} P(q_{ns}|Z_n) \prod_{t=1}^{T_n} \prod_{j \in C} P(y_{ntj}|Z_{nt}, X_{ntj}, q_{ns})^{y_{ntj}} \right)^{\frac{W_g}{H_g}}$$

## 5.  ESTIMATION RESULTS AND DISCUSSION

The destination choice model is conditional on adoption as it will be estimated from observations pertinent to adopters and users of the carsharing service. As we are assuming that individuals from the Household Travel Survey (HTS) are non-adopters, the destination choice model was estimated using data from the carsharing service and in particular the travel patterns via the carsharing service for a period of six months. The following section entails results of the destination choice model which will be used to compute the accessibility measure that is then used as an explanatory variable in the technology adoption model. Followed by that, results of the technology adoption model will be presented. Results of the destination choice model for the 16 zip code based stations are tabulated below including parameter estimates (and t-statistics).

**Table 1: Destination Choice Model**

| Variable | Parameter Estimate |
|---|---|
| Distance in 100 Kilometers | -0.24 (-2.06) |
| Employment in Zip Code (employees/miles$^2$) | 0.18 (10.06) |
| Home | 1.55 (20.51) |
| On-street Parking | 0.34 (5.47) |
| Trip between Major Technology Firm and Downtown | 1.00 (14.18) |
| Trip between Major Technology Firm and Major Airport | 2.78 (45.46) |
| <u>Alternative Specific Constant</u><br>Technology Firm | 1.10 (13.27) |
| Airport 1 | 1.76 (23.07) |
| Airport 2 | 0.61 (5.90) |
| Airport 3 | 0.93 (10.32) |

We included 4 alternative specific constants (ASCs) for 4 stations as we considered them to be hubs for trips conducted using the carsharing service. The four exogenous variables used were distance, employment rate, home dummy, and on-street parking. The on-street parking variable was introduced in the destination choice model utility specification in order to quantify and understand the effect of having on-street parking versus stations on the projected number of adopters. The on-street parking variable used was a dummy variable which will be equal to one if the destination alternative (zip code) entails on-street parking structure for the new transportation service and zero otherwise.

We did not include ASCs in all 16 utility equations because that will problematic when evaluating accessibility when new stations are introduced as it will be difficult to assess the ASC of the new destination i.e. station. In addition to that, a dummy variable between a major technology firm's headquarters and a major airport in the city was introduced which takes a value of one if a trip takes place between the technology firm and the airport stations and zero otherwise. That dummy variable was of interest as 46% of the total trips of the carsharing service had either that technology firm or airport as an origin or destination. Finally, a dummy variable between the major technology firm and the city's downtown region was introduced which takes a value of one if a trip takes place between the technology firm and downtown, and zero otherwise.

Using the parameter estimates from the destination choice model, the logsum measure of accessibility was calculated for all observations in the carsharing service data and the HTS data. The HTS data does recognize household characteristics but in our case we were only interested in the following variables: home zip code, gender and work TAZ. Finally, both datasets (HTS and carsharing service) were integrated together and used to estimate the disaggregate diffusion model.

Since we had apriori hypothesis regarding the number of latent classes in our model, determining the final model specification was based on varying the utility specification for both sub-models i.e. class membership and class-specific choice models. The table below presents detailed parameter estimates (and t-statistics) for the class membership of the technology adoption model.

**Table 2: Class Membership Model**

| Variable | Class 1 – Innovators | Class 2 - Imitators | Class 3 – Non-Adopters |
|---|---|---|---|
| Alternative Specific Constant | -- | 7.00 (37.09) | 7.51 (56.78) |
| Monthly Income ($1000) | -- | -0.23 (-13.01) | -0.04 (-3.86) |
| Male | -- | -0.77 (-8.70) | -1.72 (-23.56) |

-- Not applicable

The rho-bar-squared ($\bar{\rho}^2$) measure for this technology adoption model is almost 1.0 with a total number of 4571 individuals and 120,665 observations. $\bar{\rho}^2$ has such a high value because of the weights applied to each of the observations and the fact that the market share of the carsharing adopters is very minimal compared with the rest of the population, which forces the increase in model fit.

The class membership model includes parameter estimates which correspond to the influence of socio-demographic variables on class membership. The class membership model results reveal that all else equal, an individual is more likely to be a non-adopter, high-income groups and men are more likely to be early adopters (innovators). The monthly income used in our analysis was the average zip code based income since that socio-demographic variable was not provided in the data. Sample enumeration results denote the following split in the population across the three classes: 0.22% innovators, 16.80% imitators, and 82.98% non-adopters.

Tables 3 below presents detailed parameter estimates (and t-statistics) for the class-specific model corresponding to the adoption behavior of the new technology.

**Table 3: Class-specific Technology Adoption Model**

| Variable | Parameter | Class 1 – Innovators | Class 2 – Imitators | Class 3 – Non-Adopters |
|---|---|---|---|---|
| Alternative Specific Constant (Adoption) | $\lambda$ | -7.88 (-78.08) | -14.71 (-78.63) | -23.46 (-0.01)* |
| Major Technology Firm Employee | $\beta$ | 1.33 (1.89)* | 7.10 (46.43) | -- |
| Station in Zip Code | $\gamma$ | 1.38 (3.61) | -- | -- |
| On-street Parking in Zip Code | | 1.18 (3.99) | -- | -- |
| Accessibility for Zip Codes Containing a Station or On-street Parking | | 0.44 (5.29) | 0.68 (55.39) | -- |
| Accessibility for Zip Codes Containing neither a Station nor On-Street parking | | 0.91 (22.77) | 0.59 (22.64) | -- |
| Cumulative Number of Adopters at (t-1) in 100's | $\alpha$ | -- | 0.14 (24.21) | -- |
| Degree of Distance Friction Effect for Accessibility | $\varphi$ | 1.00 (--) | | |

-- Not applicable; * Insignificant at the 5% level

As for the class-specific model results, the parameter estimates for the utility of adoption for the two types of adopters have the right sign and are significant at the 1% level except for the major technology firm employee variable for the innovators latent class. This agrees with the behavioral interpretation of the adoption process for each class. Early Adopters' utility of adoption increases with an individual being an employee of the major technology firm and having a station or on-street parking for the new transportation service in his/her corresponding zip code.

Also, an increase in the accessibility of a certain home zip code that has neither a station nor on-street parking will in turn drive an innovator to adopt. A similar behavioral interpretation applies for home zip codes that do have stations or on-street parking. Imitators' utility of adoption increases with an individual being an employee of the major technology firm and with an increase in the cumulative number of adopters in the previous time period. This is the class which is highly influenced by previous adopters. Moreover, as the accessibility of the home zip code which has neither a station nor on-street parking increases, an imitator is more likely to adopt. The same rationale also applies for home zip codes that do have stations or on-street parking. The behavior of the non-adopters latent class is deterministic as the probability of adoption is almost equal to one for each individual that belongs to this market segment at each time period. Finally, the degree of distance friction effect for accessibility, $\varphi$, was not directly estimated as our LCCM evaluates the gradient and standard errors for linear-in-parameter utility equations. In order to account for this, we estimated several models by varying the value of $\varphi$ and selected the value that maximized the final log-likelihood. This approach is similar to a grid search.

In order to evaluate the performance of our disaggregate adoption model, we will compare its performance against a multinomial logit (MNL) model. The MNL model comprises two utility equations: adoption and non-adoption. The adoption utility specification entails all of the explanatory variables used in the LCCM while the non-adoption systematic utility specification is constrained to zero. Parameter estimates for the MNL model were behaviorally consistent and significant. The same dataset that was used to estimate model parameters for the LCCM was used for the MNL model. The final log-likelihood values for the LCCM and MNL models after accounting for choice-based sampling were -14.74, and -156.53 respectively. The table below shows the rho-bar-squared ($\bar{\rho}^2$), AIC and BIC values for the two models. It is clear that our proposed disaggregate adoption model has a slightly higher $\bar{\rho}^2$ and lower AIC and BIC values. That is why it has a better statistical fit as compared with the MNL model. It is interesting to note that both models have an extremely high $\bar{\rho}^2$ value because the marginal probability of adoption in the population is very small.

**Table 4: Measures of Model Fit**

| Model | Log-Likelihood | $\bar{\rho}^2$ | AIC | BIC |
|-------|---------------|-----------|-----|-----|
| MNL | -156.53 | 0.99 | 329 | 407 |
| LCCM | -14.74 | 1.00 | 65 | 240 |

Finally, in order to asses model performance on a hold-out sample, we estimated the parameters of the disaggregate adoption model using observations for the first 24 months. We calibrated the model to minimize discrepancy between the predicted number of adopters and the actual number of adopters for the 25th month. We then used our calibrated model to forecast adoption for months 26-30. We performed simulation using 1000 draws by bootstrapping parameter estimates of the disaggregate adoption model at each draw. Accordingly we can generate a confidence interval that bounds the predicted number of adopters. The figure below displays boxplots for the 1000 bootstrap samples in addition to the actual number of adopters for months 26-30.

It is evident that the actual number of adopters for months 27, 28 and 29 fall inside their corresponding box, which spans the first quartile to the third quartile. As for month 26, the actual number of adopters falls within the "whiskers", which is acceptable to a certain extent. This is a good indication regarding the confidence interval bound of the model's predicted number of adopters. That is definitely not the case during the 30th month as the actual number of adopters is located outside the boxplot's "whiskers". It is important to note that a new station for the carsharing service was introduced at the beginning of the 30th month. However, the actual number of adopters is much lower than previous months. That could be due to unobserved competition in the market or some issues in the carsharing service itself, which are not accounted for in our model due to limitations of the data.

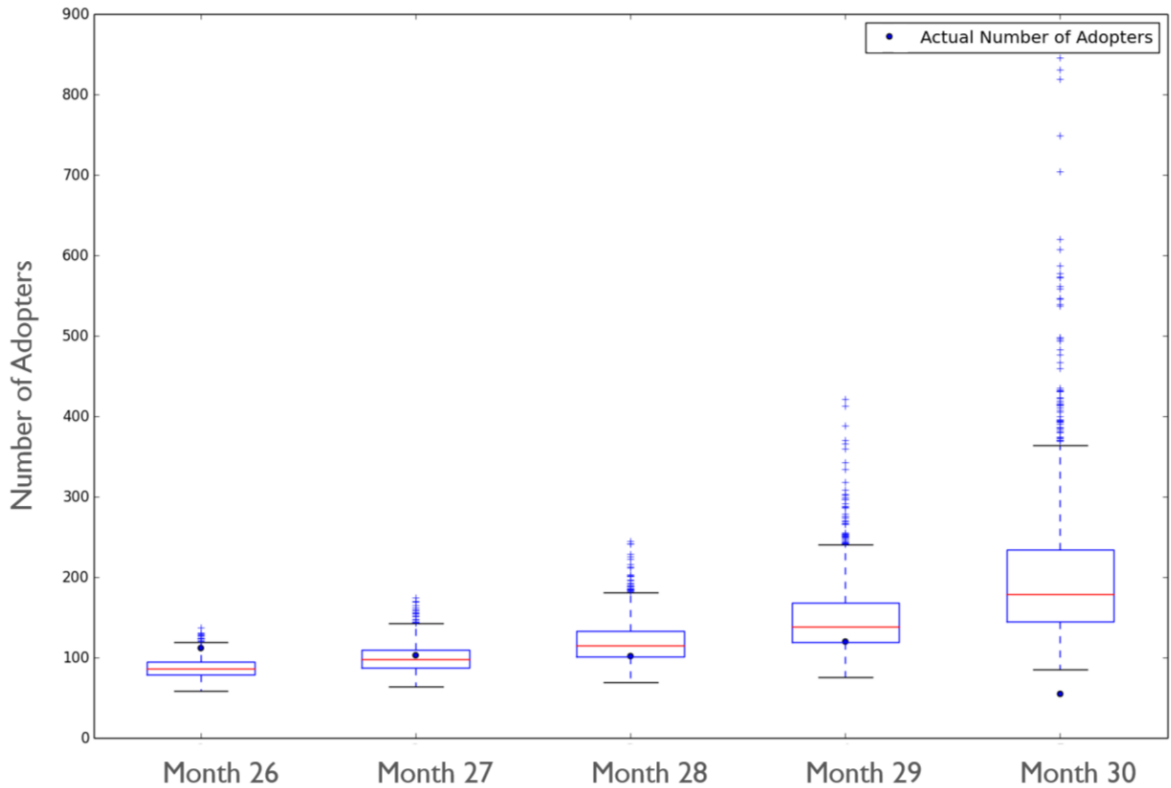

**Figure 6: Boxplots for Predicted Number of Adopters for Months 26-30**

## 6.  POLICY ANALYSIS

Now that we have estimated a technology adoption model, we want to use it to forecast adoption into the future for various potential scenarios. More specifically, we are interested in using the model to understand the potential effectiveness of new pods and on-street parking facilities placed in different locations. In order to do so, we should calibrate our model first by adjusting the values of the alternative specific constants (ASCs) of the utility of adoption for innovators and imitators. That will minimize the difference between projected and actual demand. In order to do so, we will perform sample enumeration on the entire population of the major city using our estimated model in order to predict the number of adopters that joined the service during the last month of the data's time horizon. We will adjust the ASCs in order to equate the predicted number of adopters for the last month from the model with the actual number of adopters for that month from the data itself.

There were three scenarios that we were interested in assessing their impact on the adoption of the new transportation service besides the base case scenario. The base case scenario comprises not investing in any new station or on-street parking facility in any of the zip codes. The three scenarios are:

a-  Stations/pods outside a second major technology firm
b-  Stations/pods in a new zip code in the downtown region
c-  On-street parking facilities instead of stations/pods in the same zip code as in scenario b

The figure below displays how the cumulative adoption "S" diffusion curve will be projected into the future under the aforementioned potential scenarios.

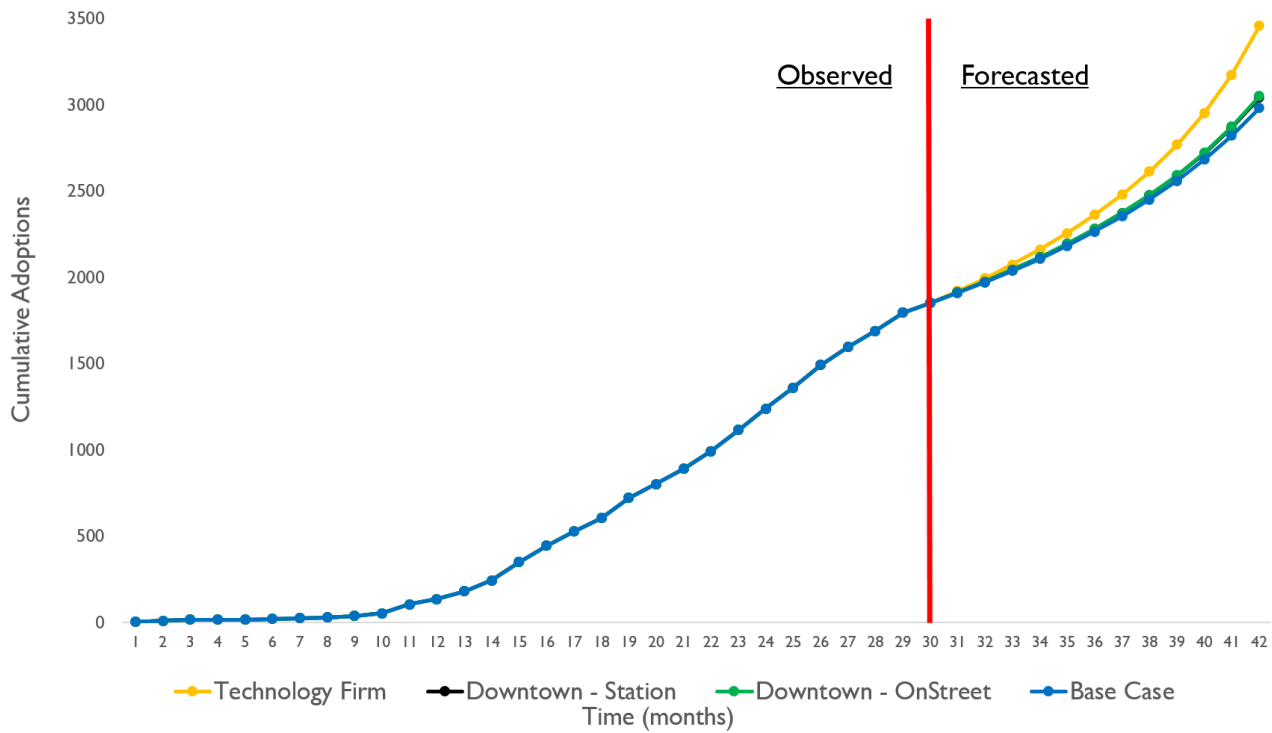

**Figure 7: Cumulative Adoptions for New Transportation Service**

Also, the figure below identifies the forecasted cumulative monthly adoptions of the new transportation service for the next year on a month to month basis. It is evident that investing in stations/pods outside another major technology firm will increase the monthly number of new adopters the most. There is no significant difference in the number of new monthly adopters for the downtown region between having a station or on-street parking. That is because, the only way we were able to incorporate the effect of each was via dummy variables. Ideally, we would have been interested in incorporating the number of cars in each station/pod or total area allocated for on-street parking but that information was not available. That said, the power of the integrated discrete choice and adoption model we developed lies in projecting adoption into the future and identifying the most effective policy that will cater for behavior change and maximize adoption.

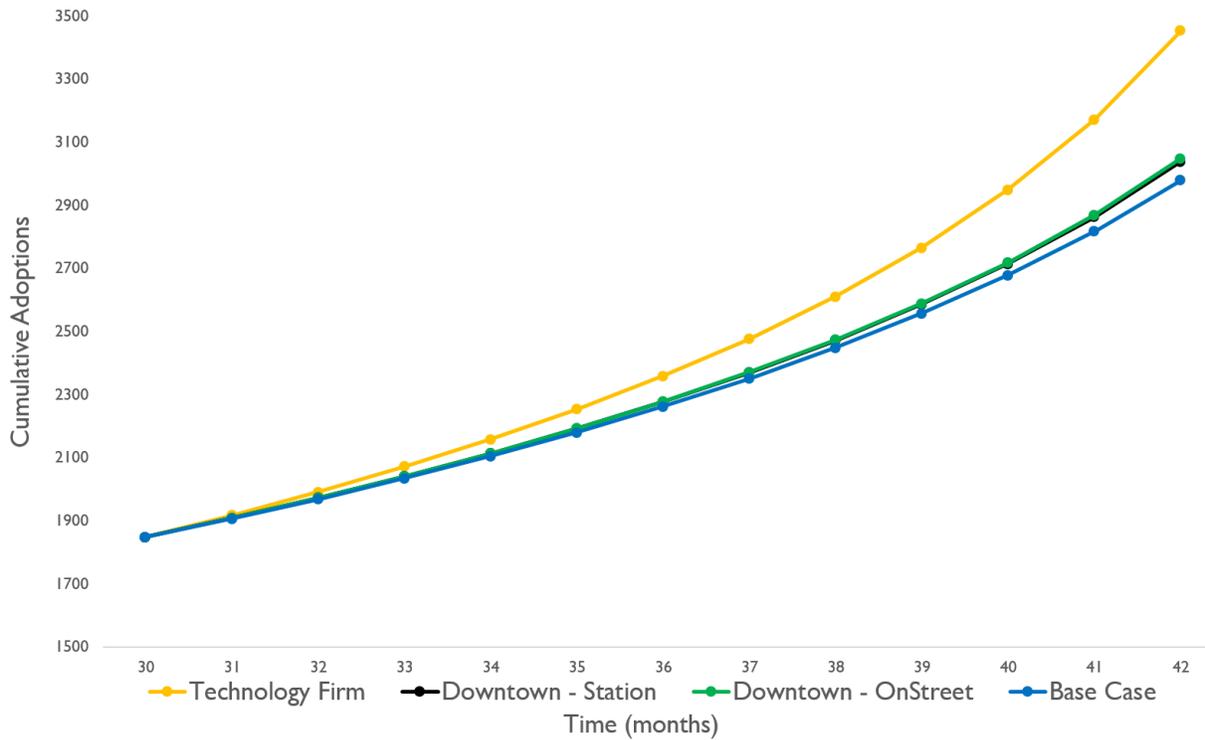

**Figure 8: Forecasted Adoption for New Transportation Service**

In order to define confidence intervals for the predicted cumulative number of adopters for each of the four aforementioned scenarios, we performed simulation using 1000 draws by bootstrapping estimates of the adoption model at each draw. The figures below display the boxplots for the 1000 bootstrap samples for the predicted cumulative number of adopters for each of the four scenarios. It is evident that the bound of the confidence interval increases with time. This indicates that predictions become more stochastic over time, which is a reasonable argument. Again, investing in stations/pods outside a major technology firm yields the highest number of forecasted adopters. Moreover, no significant difference is depicted in the predicted cumulative adopters for the downtown region for an on-street parking facility versus a station/pod.

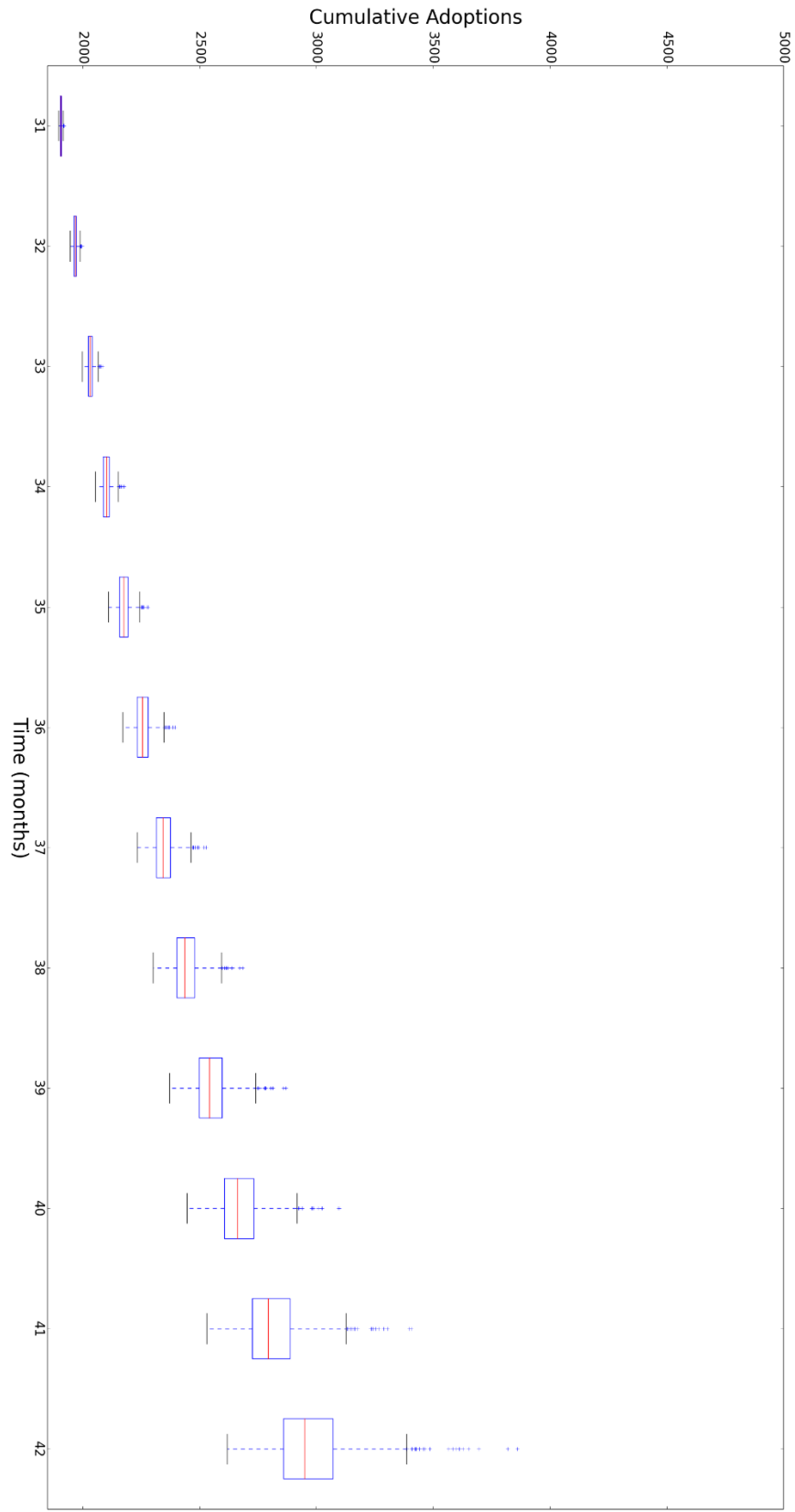

**Figure 9: Boxplots for Predicted Cumulative Number of Adopters for Base Case Scenario**

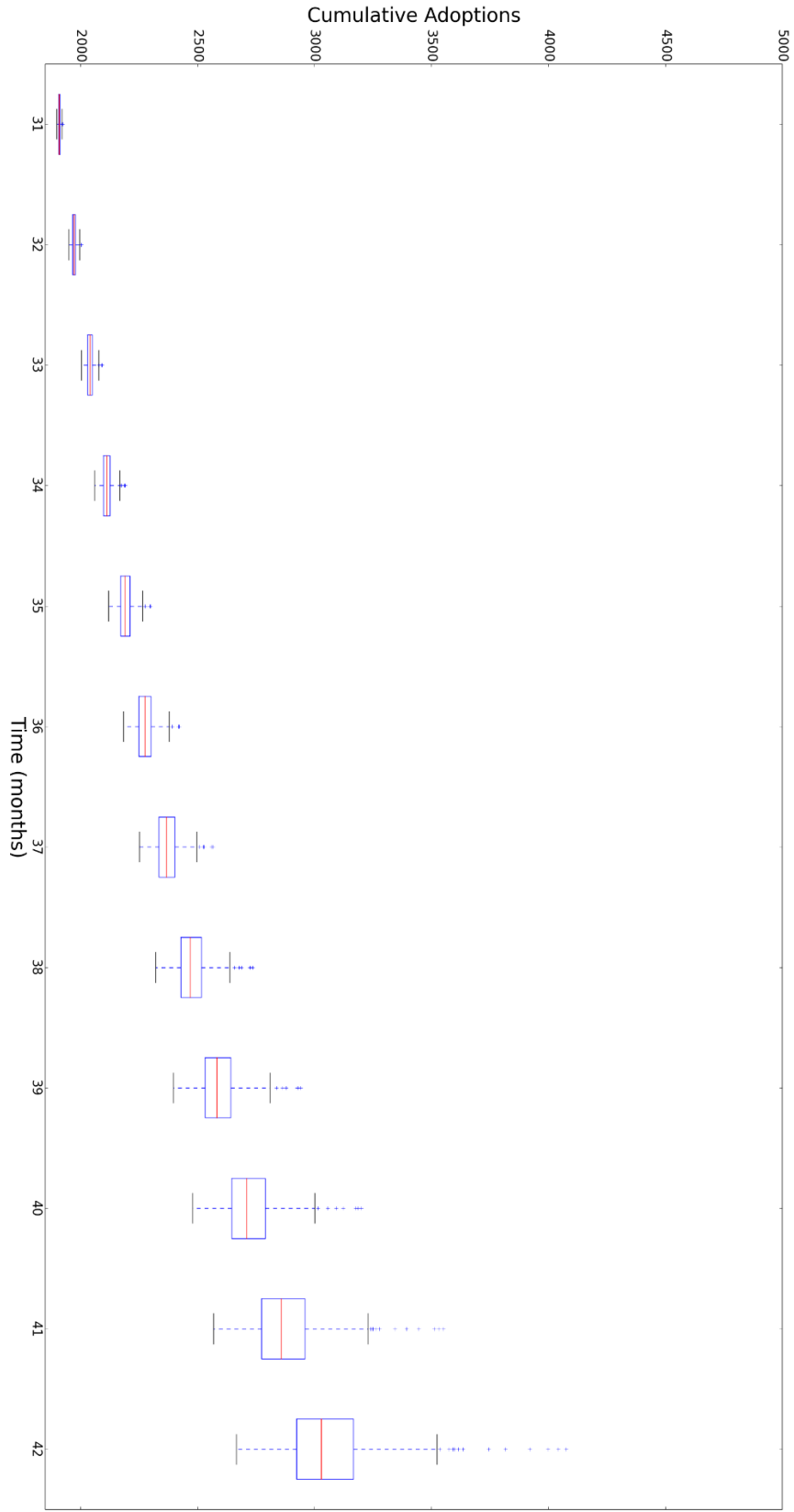

**Figure 10: Boxplots for Predicted Cumulative Number of Adopters for Downtown On-Street Parking Scenario**

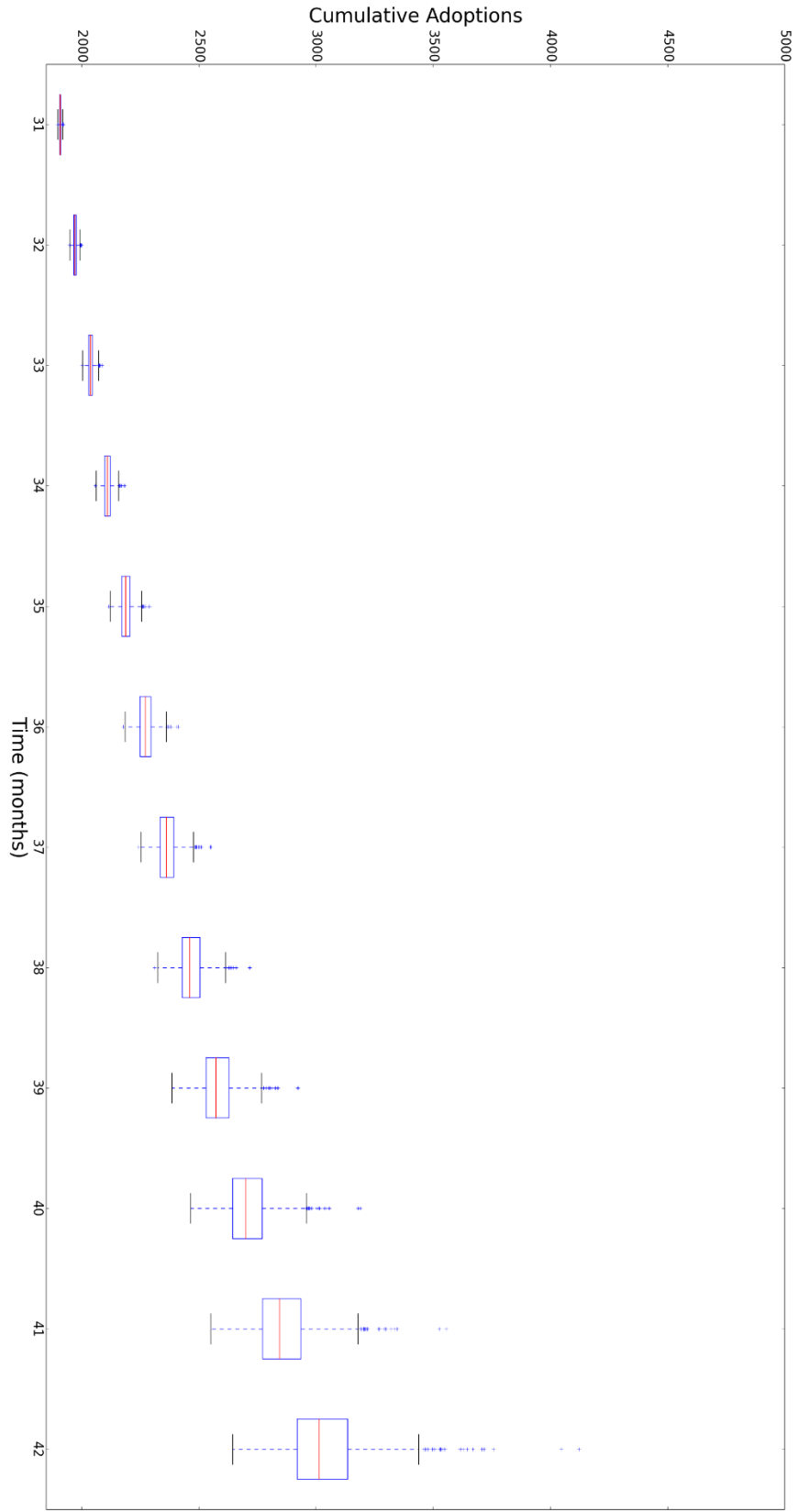

**Figure 11: Boxplots for Predicted Cumulative Number of Adopters for Downtown Station Scenario**

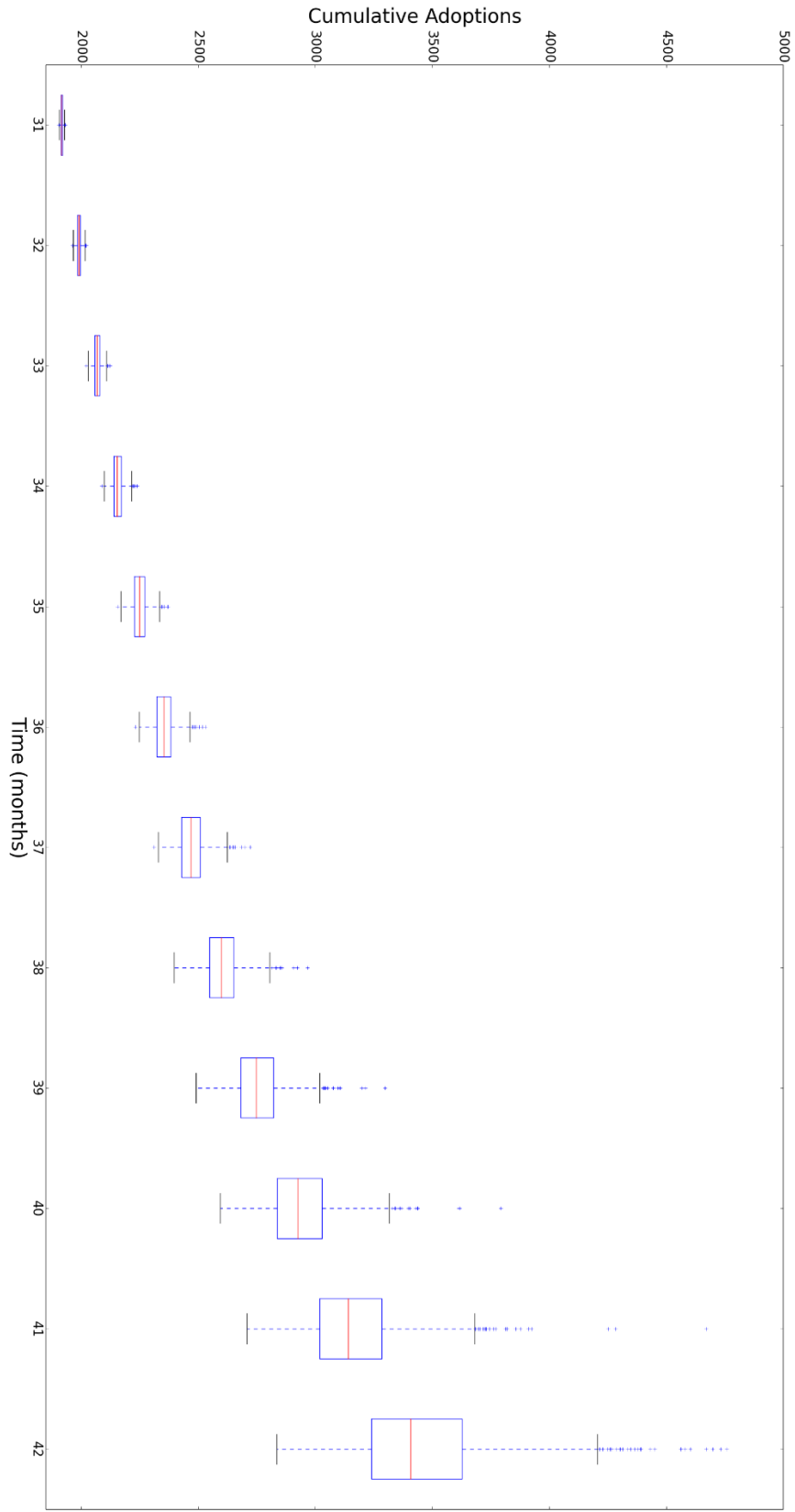

**Figure 12: Boxplots for Predicted Cumulative Number of Adopters for Major Technology Firm**

We were also interested in assessing the aggregate technology adoption process using the Bass model to highlight the advantages of our adopted methodological framework. The three variables that need to be calculated which define the "S"-shaped diffusion curve of the cumulative number of adopters of the one-way carsharing service are: the coefficient of innovation p, for coefficient of imitation q, and total potential market M. In order to compute the values for those three variables, we need to define the following formulation (Bass, 1969):

$$S(t) = pM + (q - p)Y(t) - \frac{q}{M}Y^2(t)$$

S(t) depicts sales of a product over time which is the expected number of adopters of the carsharing service at time period t. The discrete time series data was used to run the required regression analysis in order to estimate p, q and M that attained the following values respectively: 0.0051, 0.2108 and 2200. The figure below displays how the number of adopters S(t) and cumulative number of adopters Y(t) will evolve over time. The cumulative number of adopters will plateau and attain a value of 2200 adopters. This value is predicted by the Bass model and will disregard any changes in the attributes of the technology or its spatial configuration that could occur at future time periods. The Bass model suffers from the following limitations: (1) lack of including important policy variables into model parametrization which hinders its forecasting power in terms of identifying effective policies and investment strategies that maximize the expected number of adopters; (2) absence of key variables that shape the adoption process of a new transportation service such as the spatial configuration of the service; and (3) absence of incorporating the effect of socio-demographic variables onto the diffusion process which should be accounted for to capture heterogeneity in the decision making process across different consumers. That is why, the Bass diffusion model forecasts displayed below, will be identical across each of the aforementioned three potential investment strategies / policies.

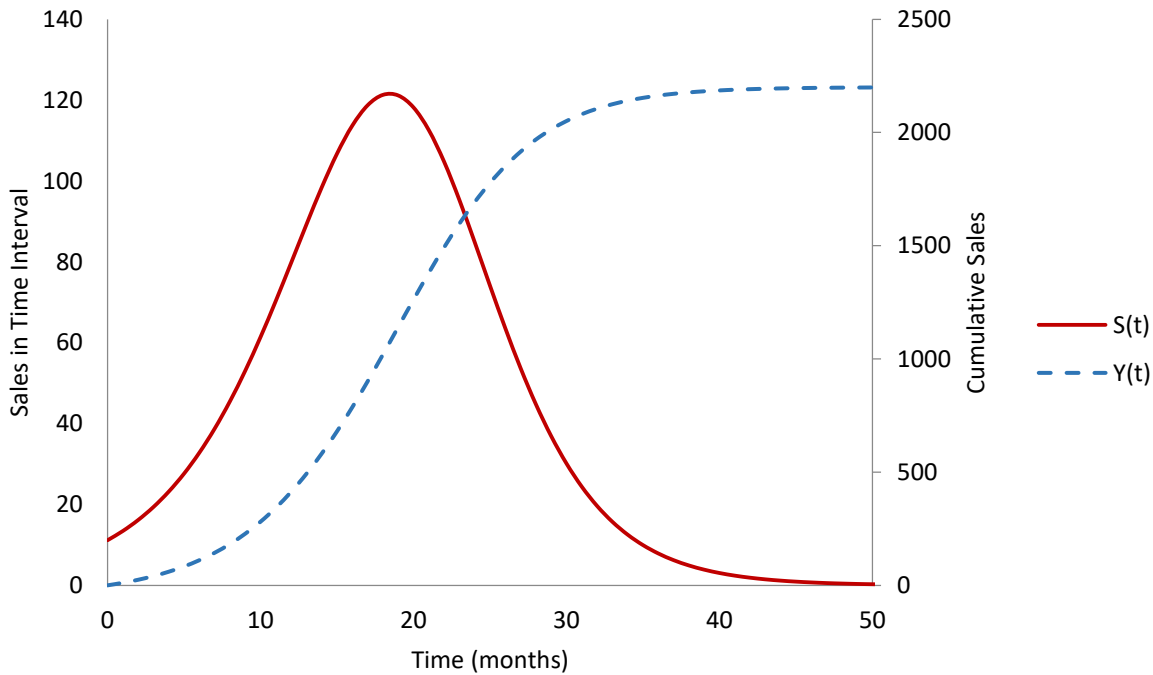

**Figure 13: Adopters vs. Cumulative Adopters over Time Using Bass Model**

# 7.  CONCLUSION

Major technological and infrastructural changes over the next decades, such as the introduction of autonomous vehicles, implementation of mileage-based fees, carsharing and ridesharing are expected to have a profound impact on lifestyles and travel behavior. However, the commonly-used approach for predicting the 20-30 year forecasts across transportation networks suffers from its inability to project membership of upcoming modes of transport. The methodological framework used in our analysis to study technology adoption consisted of an integrated latent class choice model (LCCM) and network effect model that was governed by a destination choice model. The latent classes used in the analysis are supported by the technology diffusion literature across multiple disciplines and are defined as: innovators/early adopters, imitators and non-adopters. These latent classes are able to capture heterogeneity in preferences towards technology adoption. Each class entails a distinct set of sensitivities and parameter estimates pertinent to the exogenous variables used in estimation. The adopted methodological framework focuses on understanding the relative impact of the following set of covariates: social influences, network/spatial effect, socio-demographics and level-of-service attributes.

One major contribution for this research project is defining a methodology to capture the impact of the network/spatial effect of the new technology. We were interested in understanding how the size of the network, governed by the new mode of transportation, would influence the adoption behavior of the different market segments as the ability of reaching out to multiple destination increased i.e. the size of the network grew bigger. This is a critical component in our analysis as it will quantify the effect of placing stations or on-street parking facilities in different locations and prioritize locations in the transportation network that will maximize the expected number of adopters. Our generalized technology adoption model has two other major advantages whereby it employs a microeconomic utility-maximizing representation of individuals and captures various sources of heterogeneity in the decision-making process.

The empirical results look promising in terms of defining the adoption behavior of the three classes. Finally, the model was calibrated and used to project adoption into the future for various potential scenarios. Some findings from our technology adoption model are: (1) a decision-maker is more likely to be a non-adopter, high-income groups and men are more likely to be early adopters or innovators; (2) network/spatial effect, socio-demographics, social influences and level-of-service attributes of the new technology have a positive set of sensitivities in the utility of adoption across latent classes which is consistent with our a-priori hypotheses and the diffusion literature; (3) placing a new station/pod for the carsharing system outside a major technology firm will increase the expected number of monthly adopters the most; and (4) no significant difference is observed regarding the expected number of monthly adopters for the downtown region between having a station or on-street parking.

# ACKNOWLEDGEMENTS
We would like to thank UCCONNECT (USDOT and Caltrans) for supporting our research project. We would also like to thank Susan Shaheen for her assistance in helping us get the data from the one-way carsharing system. Finally, we would like to show our gratitude to the one-way carsharing system team that was helpful in answering our questions and concerns.